\documentclass[12pt]{article}
\usepackage{amsmath}
\usepackage{epsfig}
\def\fun#1#2{\lower3.6pt\vbox{\baselineskip0pt\lineskip.9pt
\ialign{$\mathsurround=0pt#1\hfil##\hfil$\crcr#2\crcr\sim\crcr}}}

\begin{document}
\title{ Many body effects in nuclear matter QCD sum rules
}
\author{E. G. Drukarev, M. G. Ryskin, V. A. Sadovnikova\\
{\em National Research Center "Kurchatov Institute"}\\
{\em B. P. Konstantinov Petersburg Nuclear Physics Institute}\\
{\em Gatchina, St. Petersburg 188300, Russia}}
\date{}
\maketitle

\begin{abstract}
We calculate the single-particle nucleon characteristics in symmetric nuclear matter with inclusion of the $3N$ and $4N$ interactions. We calculated the contribution of the $3N$ interactions earlier, now we add that of the $4N$ ones. The contribution of the $4N$ forces to nucleon self energies is expressed in terms of the nonlocal scalar condensate ($d=3$) and of the configurations of the vector-scalar and the scalar-scalar quark condensates ($d=6$) in which two diquark operators act on two different nucleons of the matter.These four-quark condensates are obtained in the model-independent way. The density dependence of the nucleon effective mass, of the vector self energy and of the single-particle potential energy are obtained. We traced the dependence of the nucleon characteristics on the actual value of the pion-nucleon sigma term. We obtained also the nucleon characteristics in terms of the quasifree nucleons, with the noninteracting nucleons surrounded by their pion clouds as the starting point. This approach leads to strict hierarchy of the many body forces.

\end{abstract}

\section{Introduction}

The idea of QCD sum rules in vacuum is to express the characteristics of the observed hadrons in terms of the
vacuum expectation values of the QCD operators, often referred to as the condensates. Initially the method was suggested
for calculation of the meson characteristics \cite{1}. Later it was used for nucleons \cite{2}. It is described in details in \cite{3}. Later the approach was extended for the case of finite baryon density (see the  review \cite{4} for references). The idea of QCD sum rules in nuclear matter is
to express the self energies of  {\em the probe proton} in terms of the in-medium condensates.

 We  gave a  short review of the  approach in Introduction to our recent paper \cite{5}, and there is no need to repeat it here. The self energies of {\em the probe proton} are expressed in terms of the in-medium condensate.
We just remind the readers that
the sum rules are based on the dispersion relations for the function $\Pi(q)$ which carries the proton quantum numbers. The dispersion relations are considered at large and negative values of $q^2$. Here the function $\Pi(q)$ can be calculated as a power series of $q^{-2}$ with QCD condensates as the coefficients of the expansion (the Operator Product Expansion, OPE). The imaginary part of $\Pi(q)$ on the right hand side of the dispersion relations is formed by the real physical states with the quantum numbers of the proton. Here we separate the lowest lying pole corresponding to the probe proton. In nuclear matter we introduce  the 4-vector $P=(m,{\bf 0})$ with $m$ being the vacuum value of the nucleon mass (we neglect the neutron-proton mass splitting). There are three structures of the function $\Pi(q)$ proportional to the matrices $\hat q$, $\hat P$ and to the unit matrix $I$. The Borel transformed sum rules take the form
\begin{equation}
{\cal L}^q(M^2, W_m^2)=\Lambda_m; \quad {\cal L}^I(M^2, W_m^2)=m^{*}\Lambda_m; \quad {\cal L}^P(M^2, W_m^2)=-\frac{\Sigma_V}{m}\Lambda_m.
\label{X1}
\end{equation}
Here the left hand sides are the Borel transformed OPE terms (the Borel transform is labeled by the tilde sign). They can be presented as
\begin{equation}
{\cal L}^q(M^2,W_m^2)=\sum_{n=0}\tilde A_n(M^2,W_m^2); \quad {\cal L}^I(M^2,W_m^2)=\sum_{n=3}\tilde B_n(M^2,W_m^2);
\label{X2}
\end{equation}
$${\cal L}^P(M^2,W_m^2)= \sum_{n=3}\tilde C_n(M^2,W_m^2).$$
Here $n$ stands for the dimension of the condensates.
The right hand sides of Eq.(\ref{X1}) contain the effective mass of the probe proton $m^*$ and its vector self energy $\Sigma_V$.
They are multiplied by the factor
$\Lambda_m=\lambda^2_{m}e^{-m_m^2/M^2}$ with $\lambda_m^2=32\pi^4\lambda_{Nm}^2$ while $\lambda_{Nm}$ is the residue of the nucleon pole. The position of the pole $m_m$ is expressed in terms of the characteristics of the probe nucleon $m^*$ and $\Sigma_V$.

The condensates of the lowest dimension $d=3$ are the vector and scalar quark expectation values
$v(\rho)=\langle M|\sum_i\bar q^i\gamma_0q^i|M\rangle$ and
$\kappa(\rho)=\langle M|\sum_i\bar q^iq^i|M\rangle.$
Here $|M\rangle$ is the ground state of nuclear matter, the sums are carried out over the quark flavors,
$\rho$ is the nuclear matter density. The gluon condensate and the nonlocal contributions to the vector condensate have dimension $d=4$.
They are included in the calculations, providing the contributions which are numerically small. Thus we do not discuss them here.
The contributions of dimension $d=6$ are provided mostly by the four quark condensates
 $\langle M|q^a_{\alpha}\bar q^b_{\beta}q^c_{\gamma}\bar q^d_{\delta}|M\rangle$.
The vector condensate $v(\rho)=n_v\rho$, where $n_v=3$ is the number of the valence quarks in nucleon, is exactly proportional to the nucleon density. The density behavior of the other condensates depends on the approximations used for description of the nuclear matter.

It is reasonable to start the analysis treating the nuclear matter as
a relativistic Fermi gas  of free noninteracting nucleons. We call this a gas approximation. In such approximation each condensate can be presented as the sum of the vacuum value and nucleon density times expectation value of the same operator in the free nucleon. Neglecting the Fermi motion of the nucleons we can write
\begin{equation}
\kappa(\rho)=\kappa(0)+\kappa_N\rho,
\label{3o}
\end{equation}
with the nucleon matrix element for the nucleon at rest
\begin{equation}
\kappa_N=\langle N|\sum_i\bar q^i(0)q^i(0)|N\rangle.
 \label{3}
\end{equation}
It can be expressed through the observable pion-nucleon sigma term $\sigma_N$ \cite{10}
\begin{equation}
\kappa_N=\frac{2\sigma_N}{m_u+m_d},
\label{3z}
\end{equation}
with $m_{u,d}$ the masses of light quarks. The sigma term can be expressed as
\begin{equation}
\sigma_N=m_{\pi}^2\frac{\partial E^{(0)}}{\partial m^2_{\pi}}.
\label{3a}
\end{equation}
Here $m_{\pi}$ is the pion mass while $E^{(0)}=m$ is the energy of free nucleon at rest.

For the nucleon with three momentum ${\bf p}$ one can find the corrections of the order $p^2/m^2$ to the right hand side of Eq.(\ref{3}), replacing $E^{(0)}$ by $E=\sqrt{m^2+p^2}$ on the right hand side of Eq.(\ref{3a}). After integration over the Fermi sphere of the radius $p_F$ one finds \cite{10a}
\begin{equation}
\kappa(\rho)=\kappa(0)+\kappa_Nt(p_F)\rho ; \quad t(p_F)=1+\frac{3p_F^2}{10m^2}-\frac{3p_F^4}{56m^4} + O(\frac{p_F^6}{m^6}).
 \label{3b}
\end{equation}
Recall that $\rho=2p_F^3/3\pi^2$. If the kinematical corrections are neglected, i.e. $t(p_F)=1$ the scalar condensate $\kappa(\rho)$ is expressed by Eq.(\ref{3o}) and is linear in $\rho$.

The gas approximation corresponds to interaction of the probe proton with each of the nucleons of the matter separately. Hence Eq.(\ref{X1}) provides the nucleon characteristics with inclusion of the $2N$ forces only \cite{4}.

Contribution of the $3N$ forces to the self energies of the probe proton was found in  \cite{5} by inclusion of the in-medium QCD condensates beyond the gas approximation and taking into account only the  $2N$ forces between the nucleons of the matter. Neglecting the kinematical corrections we can write
\begin{equation}
E=m+ E^{(2)}
\label{3a16}
\end{equation}
with $E^{(2)}$ the energy of interactions between the nucleons of the matter with inclusion of only the $2N$ forces.
Define similar to Eq.(\ref{3a})
\begin{equation}
\sigma^{(2)}(\rho)=m_{\pi}^2\frac{\partial E^{(2)}}{\partial m^2_{\pi}},
\label{3a1}
\end{equation}
and
\begin{equation}
\kappa^{(2)}(\rho)=\frac{2\sigma^{(2)}(\rho)}{m_u+m_d}.
 \label{3a2}
\end{equation}
The scalar condensate is thus
\begin{equation}
\kappa(\rho)=\kappa_0+(\kappa_N+\kappa^{(2)}(\rho))\rho.
\label{4a0}
\end{equation}

Here we neglected corrections caused by the nucleon kinetic energy. Note that the latter are not as simple as in the gas approximation (Eq.(\ref{3b})). Indeed, the nucleons of the matter can be viewed as moving in certain vector and scalar fields $E_V$ and $E_S$ with the energy $E=E_V+\sqrt{(m+E_S)^2+p^2}$. Thus expansion in powers of $p^2$ should be rather that in powers of $p^2/(m+E_S)^2$.
At saturation value of nuclear density $E_S \approx -0.4 m$ \cite{12a}, and thus $E_S$ can not be neglected.
Note also that $E_S$ depends on $m_{\pi}^2$.

 There is also a contribution to the $3N$ forces acting on the probe proton, originated by configurations of the four-quark condensates in which two pairs of quarks act on two different nucleons of the matter contribute to the $3N$ forces.

Now we want to calculate the proton self energies with inclusion of the $4N$ forces. The nucleon energy is
\begin{equation}
E=m+ E^{(2)}+E^{(3)}
\label{3a19}
\end{equation}
with $E^{(3)}$ the contribution of the $3N$ forces to the energy of interactions between the nucleons of the matter.
Introducing similar to Eq.(\ref{3a1})
\begin{equation}
\sigma^{(3)}(\rho)=m_{\pi}^2\frac{\partial E^{(3)}}{\partial m^2_{\pi}}; \quad \kappa^{(3)}(\rho)=\frac{2\sigma^{(3)}(\rho)}{m_u+m_d},
\label{4a1}
\end{equation}
we define
$$\sigma^{eff}(\rho)=\sigma_N+\sigma^{(2)}(\rho)+\sigma^{(3)}(\rho),$$
The quark condensate becomes
\begin{equation}
\kappa(\rho)=\kappa_0+\kappa^{eff}(\rho)\rho; \quad \kappa^{eff}(\rho)=\kappa_N+\kappa^{(2)}(\rho)+\kappa^{(3)}(\rho).
\label{4a2}
\end{equation}
One can view $\kappa^{eff}$ as the value of the operator $\sum_i\bar q^i(0)q^i(0)$ averaged over the nucleon bound in nuclear matter with the $2N$ and $3N$ forces included.

In the meson exchange picture the nucleons of the matter exchange by the mesons $\mu_i$. The contributions $\sigma^{(n)}=d E^{(n)}/dm_q$ with $m_q=(m_u+m_d)/2$ should be summed over the mesons $\mu_i$. The contribution of each meson to $\sigma^{(n)}$ is proportional to the expectation value
$w_i=\langle \mu_i|\sum_i\bar q^i(0)q^i(0)|\mu_i\rangle$ which is (under the proper normalization) just the total number of  quarks and antiquarks in the meson. All the mesons, but the pions contain mainly the valence quarks. Thus the simple quark counting provides $w_i \approx 2$. The pions contain the sea of the quark-antiquark pairs, and one can calculate $w_{\pi}=m_{\pi}/(m_u+m_d) \approx 12$. Thus we expect the pion exchanges to give the leading contribution to $E^{(n)}$. Only these contributions will be included.

The configurations in which two pairs of quark operators act on two different free nucleons contribute to the $2N$ forces between  the nucleons of the matter. Thus they contribute to the $3N$ forces acting on the probe proton. The contributions of the vector- scalar
($VS$) and of the scalar-scalar ($SS$) four-quark condensates which can be obtained in the model-independent way \cite{5} depend on the nucleon matrix element
$\kappa_N$. Including the lowest in-medium corrections to $\kappa_N$ we find the contribution of the $3N$ forces to the four-quark condensates leading to the $4N$ interactions between the probe nucleon an nuclear matter. The $VS$ condensate is proportional to $\kappa_N$. To find the contribution of the $4N$ interactions to the parameters of the probe nucleon one should change $\kappa_N$ by $\kappa^{(2)}$ in the expression for the $VS$ condensate \cite{5}. In the same way the $SS$ condensate containers two factors
 $\kappa_N$. To obtain the contribution of the $4N$ interactions one should change one of the matrix elements $\kappa_N$ by $\kappa^{(2)}$ multiplying the result by the factor $2$ reflecting the permutations.

  We can establish connection between our approach and the picture based on the nucleon-nucleon interactions. The lowest dimension OPE terms are determined by exchange of weakly correlated quarks with nuclear matter. These terms determine the probe proton self energies $\Sigma_S=m^*-m$ and $\Sigma_V$. On the other hand these parameters can be considered as caused by the exchanges of the systems of strongly correlated quarks (the mesons) between the probe nucleon and nuclear matter. Thus the exchange by systems of strongly correlated quarks between the probe nucleon and the nucleons of the matter is expressed through those of weakly correlated quarks with the same quantum numbers between the three quark system and the matter. The Feynman diagrams of nucleon-nucleon interactions corresponding to inclusion of $2N$ and $3N$ forces were presented in \cite{5}.
 Additional diagrams for the $4N$ forces are given in Fig.1.

 To analyze the role of various contributions it is convenient to present
\begin{equation}
\frac{{\cal L}^I(M^2, W_m^2)}{{\cal L}^q(M^2, W_m^2)}=m^{*};\quad
\frac{{\cal L}^P(M^2, W_m^2)}{{\cal L}^q(M^2, W_m^2)}=-\Sigma_V/m\,.
\label{10}
\end{equation}
These equations follow immediately from Eq.(\ref{X1}).
The two equations become independent if we put $W_m^2=W_0^2$ with $W_0^2$ the vacuum value of the continuum threshold.
Under this assumption Eqs.(\ref{10}) express the nucleon parameters in terms of condensates.
In the total solution the unknown $W_m^2$ ties the equations.

The contributions of the $3N$ forces to the vector self energy $\Sigma_V$ manifest themselves in terms of the four-quark condensates with two pairs of quark operators acting on two different nucleons of the matter located in the same space -time point. The probability of such configuration is small due to strong repulsion of the two nucleons, known as the "core" in traditional nuclear physics. It is known to be due mostly to the Pauli principle for the quarks in the six-quark system \cite{6}. The result was supported by the lattice QCD stimulations \cite{7}, \cite{8}.
In our calculations the form of the current $\Pi(q)$ insures the needed antisymmetrization and leads to numerically small value of the contribution. Still smaller contribution comes from $4N$ forces since $\sigma^{(2)}(\rho)$ is much smaller than the linear contribution $\kappa_N\rho$ to the scalar condensate (see, e.g. \cite{4}). Thus the many body contributions to $\Sigma_V$ exhibit the standard hierarchy with domination of the gas approximation term while the $3N$ forces provide a small correction and inclusion of $4N$ forces lead to a still smaller correction.

The contribution of the $3N$ and $4N$ forces to the effective mass $m^*$ manifest themselves mainly through the nonlinear scalar condensates (NSC) $\kappa^{(2)}(\rho)\rho$ and $\kappa^{(3)}(\rho)\rho$ correspondingly. These condensates have been obtained in
\cite{9} in framework of the Chiral Perturbation Theory (CHPT). Still the NSC is much smaller than the linear part of the condensate and the contribution of $3N$ and $4N$ forces is much smaller than that of $2N$ forces. However it follows from the equations presented in \cite{9} that due to some cancelations the contribution $\sigma^{(2)}$ is several times smaller than  $|\sigma^{(3)}|$ ($\sigma^{(3)}<0$). Thus the contribution of $4N$ forces to the effective mass $m^*$ appears to be larger than that of $3N$ forces.

The contributions of the $3N$ and $4N$ forces to the vector self energy $\Sigma_V$ manifest themselves in terms of  the four-quark condensates. Due to small value of $\sigma^{(2)}$ the contribution of the $4N$ forces to $\Sigma_V$ is much smaller than that of the $3N$ forces.

We employ the results of \cite{9} for NSC with minor modifications which we discuss in the next Section. We found the density dependence of nucleon parameters with inclusion of the many body effects. The main effect of many body forces is the change of the value of nucleon effective mass $m^*$. At $\rho=\rho_0$ ( $\rho_0=0.17 fm^{-3}$ is the phenomenological value of saturation density) and $\sigma_N=45$ MeV the $3N$ forces diminish $m^*$ by $26$ MeV. The $4N$ forces increase $m^*$ by $73$ MeV. Recall that the $3N$ forces diminish $\Sigma_V$ by 37 MeV.
The $4N$ forces change $\Sigma_V$ by several MeV.

Note that the calculations of [12] include the contributions provided by both the direct (Hartree) and the exchange (Fock) interactions between the nucleons of the matter. We do not include the exchanges between our three-quark system described by the function $\Pi(q)$ and the matter since such terms correspond to the higher order terms of the operator expansion (see, e.g. \cite{4}). Thus in our approach the exchange terms of interaction between the
probe nucleon as a whole and the matter are expected to be small and are not included.

The linear part of the condensate and the NSC, as well as of the scalar-scalar and vector-scalar four-quark condensates depend on the value of nucleon sigma term $\sigma_N$. While the conventional value is $\sigma_N=(45\pm 8)$MeV \cite{10}, one can meet the larger values of $\sigma_N$ in literature (see, e.g. \cite{11}). The value $\sigma_N=(66\pm 6)$MeV \cite{12} is the largest one. We traced dependence of nuclear parameters on the value of $\sigma_N$.

Note that definition of many body effects is ambiguous. We treated the single pion exchange as a two-nucleon interaction. On the other hand it can be expressed as the nucleon self energy in the matter, which can be treated as a single particle characteristics of a quasifree nucleon. The same refers to iterated one pion exchange \cite{13}. The nucleon energy with inclusion of $1\pi$ and iterated $1\pi$ exchanges can be written as $E_{N\pi}=m+E_{N\pi}^{(2)}+E_{N\pi}^{(3)}$ with $E_{N\pi}^{(2,3)}$ corresponding to inclusion of $2N$ and $3N$ interactions.
In the gas approximation for the quasifree nucleons we can write similar to Eq.(\ref{3a})
\begin{equation}
\hat\sigma_N(\rho)=m_{\pi}^2\frac{\partial E_{N\pi}}{\partial m^2_{\pi}}.
\label{10a}
\end{equation}
 Beyond the gas approximation and with inclusion of the $2N$ and $3N$ forces we can write for the energy of nucleon interaction in the quasifree nucleon presentation
\begin{equation}
E=E_{N\pi}+\hat E^{(2)}+\hat E^{(3)}; \quad \hat E^{(n)}=E^{(n)}-E^{(n)}_{N\pi}; \quad n=2,3,
\label{10u}
\end{equation}
and define
\begin{equation}
\hat\sigma^{(n)}(\rho)=m_{\pi}^2\frac{\partial\hat E^{(n)}}{\partial m^2_{\pi}}; \quad n=2,3.
\label{10w}
\end{equation}
We can solve the sum rules in the quasifree nucleon presentation employing the equations presented above in which
$\sigma_N$ and $\sigma^{(2,3)}$ are replaced by $\hat\sigma_N$ and $\hat \sigma^{(2,3)}$ correspondingly.

In Sec.2 we calculate the contribution of the nonlinear scalar condensate. In Sec.2 we find the condensates contributing to the $4N$ forces. We obtain the density dependence of nucleon parameters with inclusion of the $3N$ and $4N$ forces in Sec.3. In Sec. 4 we solve the sum rules in the quasifree nucleon presentation. We summarize in Sec.5.

\section{Condensates contributing beyond the gas approximation}
\subsection{Nonlinear scalar condensate}

The scalar condensate contributes only to equation for ${\cal L}^I$ in Eq.(\ref{X1}). It provides \cite{4}
\begin{equation}
\tilde B_3=-4\pi^2M^4E_1(W_m^2/M^2 )\kappa(\rho).
\label{6}
\end{equation}
Note that contributions of the higher real states to the right hand side of the dispersion relation
are approximated by continuum with the effective threshold $W_m^2$ and are accounted for by the function
$E_1(x)=1-(1+x)e^{-x}$
with $x=W_m^2/M^2$. We do not write down the lowest order radiative corrections. However they are included in the calculations.

To calculate the condensate $\kappa(\rho)$ we need to know the contributions $\kappa^{(2)}$ and $\kappa^{(3)}$ expressed through
the terms $\sigma^{(2)}$ and $\sigma^{(3)}$ by Eqs.(\ref{3a2}) and (\ref{4a1}). The contribution $\kappa^{(2)}$ was obtained in \cite{5}.
Now we obtain $\kappa^{(3)}$.

It includes the iterated one-pion exchange terms involving three in-medium nucleons, expressed by Eqs.(8) and (9) of \cite{9}.
We must also include three-body contributions to the two-pion exchange terms expressed by Eqs.(12) and (13) of \cite{9}.
There are specific for the CHPT $\pi \pi NN$ vertices proportional to $c_1$. The three-body terms proportional to $c_1$ are given by
Eqs.(22) and (23) of \cite{9}. These terms provide the contribution
\begin{equation}
\sigma^{(3)}(\rho)=m_{\pi}^2\varphi(\rho)
\label{60}
\end{equation}
to the effective sigma term. Here $\varphi(\rho)$ is the sum of the right hand sides of Eqs.(8), (9), (12), (13),(22) and (23) of \cite{9}.

There is additional contribution.
Note that in the lowest order of CHPT
$$c_1=-\frac{1}{4}\frac{\partial E}{\partial m_{\pi}^2},$$
with $E$ standing for the energy of the in-medium nucleon. On the other hand, the effective nucleon sigma term
$$\sigma^{eff}=m_{\pi}^2\frac{\partial E}{\partial m_{\pi}^2},$$
 Thus in the lowest order of
CHPT $c_1=-\sigma^{eff}/4$.
The higher order CHPT terms are included by Eq.(19) of \cite{9} which can be written as
\begin{equation}
c_1=\frac{\sigma^{eff}/m_{\pi}^2+A}{-4+3m_{\pi}^2\ln{(m_{\pi}/{\mu})}/(2\pi^2f_{\pi}^2)},
\label{7}
\end{equation}
with $\mu$ the high momentum cutoff while $A$ is proportional to $g_A^2m_{\pi}/f_{\pi}^2$ and does not depend on $\sigma^{eff}$.
The explicit form of the term $A$ can be easily deduced from Eq.(19) of \cite{9}.

One must put the vacuum value of $c_1$ with $\sigma^{eff}=\sigma_N$ in Eqs.(22) and (23) of \cite{9} for the three body contributions.
However one must include also the lowest order in-medium correction to $c_1$
\begin{equation}
\delta c_1=\frac{\sigma^{(2)}/m_{\pi}^2}{-4+3m_{\pi}^2\ln{(m_{\pi}/{\mu})}/(2\pi^2f_{\pi}^2)}.
\label {7b}
\end{equation}
Combined with  Eq.(20) of \cite{9} it provides the contribution to $\sigma^{(3)}$ which is
\begin{equation}
\delta \sigma^{(3)}=\frac{3g_A^2m_{\pi}^4\sigma^{(2)}}{280\pi^3f_{\pi}^4}\cdot\frac{F(u)}{-4+3m_{\pi}^2\ln{(m_{\pi}/{\mu})}/(2\pi^2f_{\pi}^2)},
\label{7a}
\end{equation}
with $u=p_F/m_{\pi}$, while
$$F(u)=(14u^2+3u^4)\arctan {u}+\frac{27+49u^2}{4u^3}\ln{(1+u^2)}-\frac{27}{4u}-\frac{71u}{8}-\frac{99u^3}{2}.$$

\subsection{Four-quark condensates}

Going beyond the gas approximation we must include the contribution to the four quark condensate $\langle M|\bar q\Gamma_Aq\bar q\Gamma_Bq|M\rangle$ in which two pairs of quarks act on two different nucleons of the matter, which are at the same space-time point. As we said in Introduction, the strong short range repulsion of the two nucleons is due mostly to the Pauli principle for the quarks in the six-quark system \cite{6}-\cite{8}.
In the present calculations, as well as it was in those carried out in \cite{5} the form of the function $\Pi(q)$ insures the needed antisymmetrization, and the latter effectively takes into account  the main part of the nucleon short range interaction.

In \cite{5} the contribution to $\sigma^{(2)}$ was expressed in terms of the expectation values in the free nucleons.
 The contribution of
the vector-scalar condensate was thus proportional to expectation value $\kappa_N$ determined by Eq.(\ref{3}). The contribution of
the scalar-scalar condensate was proportional to $\kappa_N^2$.
To include the $3N$ forces in the matter we must replace $\kappa_N$ by $\kappa_N+\kappa^{(2)}$ in the vector-scalar condensate and $\kappa^2$ by $\kappa_N^2+2\kappa_N\kappa^{(2)}$ in the scalar condensate. The term $(\kappa^{(2)})^2$ would correspond to the $4N$ forces in nuclear matter and should be omitted.

The vector-scalar condensate is determined by the expectation value of $2u2d$ operator. Employing Eq.(36) of \cite {5}
we find that it  contributes only to the scalar structure ${\cal L}^I$ with
\begin{equation}
\tilde B_6^{4N}=18\pi^4m\kappa ^{(2)}(\rho)\rho^2.
\label{8}
\end{equation}

The expectation value of $4u$ operator provides the scalar-scalar condensate contributing to  ${\cal L}^q$.
Employing Eq.(33) of \cite{5} we find
\begin{equation}
\tilde A_6^{4N}=4\pi^4\kappa_N\kappa ^{(2)}(\rho)\rho^2.
\label{9}
\end{equation}

\section{Contribution of $3N$ and $4N$ forces to the values of nucleon parameters}

\subsection{Contribution of $4N$ forces}
Contribution of $3N$ forces was studied in \cite{5}. Now we study what do the $4N$ forces add.

Putting $W_m^2=W_0^2$ in Eq.(\ref{10}) we see that
the $4N$ forces caused by the nonlinear scalar condensate (NSC) and the scalar-scalar ($SS$) four quark condensate contribute only to the effective mass.
Since the vector-scalar ($VS$) four quark condensate contributes to the structure ${\cal L}^q$, the corresponding $4N$ forces influence both the effective mass $m^*$ and the
vector self energy $\Sigma_V$. In the total solution the unknown $W_m^2$ ties the equations. Variations of the NSC and SS condensates provide also minor changes in the value of $\Sigma_V$.

Now we solve Eqs. (\ref{X1}) including the $4N$ forces. The terms corresponding to $2N$ and $3N$ forces are given in \cite{4} and \cite{5}.
We add the contributions corresponding the $4N$ forces described in previous Section.
The contribution of $4N$ forces to the effective mass $m^*$ is determined by the contribution $\sigma^{(3)}$ to the NSC.
We found that $\sigma^{(3)}=-9.49$ MeV at $\rho=\rho_0$ while $\sigma^{(2)}=2.83$ MeV at this point \cite{5}.
Thus the contribution of the $4N$ forces to the effective mass is several times larger that of the $3N$ forces.
Note that $\sigma^{(3)}$ is dominated by the Hartree term of the iterated one-pion exchange  (Eq.(8) of \cite{12}).
The contribution $\delta \sigma^{(3)}$ determined by Eq.(\ref{7a}), which makes $0.42$ MeV at $\rho=\rho_0$ is not very important numerically but
is needed for the consistency of the procedure. Thus at $\rho=\rho_0$ the $4N$ forces increase the nucleon effective mass $m^*$ by $73$ MeV in agreement with estimations made in \cite{5}.

The density dependence of nuclear parameters with and without inclusion of the $4N$ forces is presented in Figs.2,3. More detailed results in the vicinity of the density $\rho_0$ are given in Table ~1.

\begin{table}
\caption{Contribution of the $3N$ and $4N$ forces to the density dependence of nucleon parameters. For each value of $\rho/\rho_0$
the upper line shows the result in the gas approximation (only  the $2N$ forces are included). Second line-$3N$ forces are added; third line-$3N$ and $4N$ forces are added ($\mu=882$ MeV).}
\begin{center}
\begin{tabular}{|c|c|c|c|c|c|}
\hline
$\rho/\rho_0$&$m^*$,MeV&$\Sigma_V$, MeV&$U$, MeV&$\lambda_m^2$,GeV$^6$&$W_m^2$, GeV$^2$\\
\hline
0.90&640&196&-93&1.43&1.75\\
&618&166&--145&1.44&1.79\\
&675&173&-80&1.62&1.88\\
\hline
0.95&620&209&-100&1.39&1.73\\
&596&176&-157&1.41&1.79\\
&661&183&-85&1.61&1.88\\
\hline
1.00&599&222&-108&1.36&1.72\\
&574&185&-169&1.39&1.79\\
&647&192&-89&1.59&1.88\\
\hline
1.05&577&236&-115&1.32&1.71\\
&551&195&-183&1.37&1.79\\
&632&202&-93&1.58&1.88\\
\hline
1.10&555&250&-123&1.29&1.70\\
&528&204&-196&1.36&1.80\\
&618&212&-98&1.57&1.88\\
\hline

\end{tabular}
\end{center}
\end{table}

The contribution of the $4N$ forces can be seen by comparing the solid and dashed lines in Figs.2,3 and of the
numbers in the third and second lines of Table 1.

As one can see from Table 1, the $4N$ forces also increase the vector self energy $\Sigma_V$ by $7$ MeV. Note, however that the result for $\Sigma_V$ is determined by the shift of the continuum threshold value $W_m^2$ in the self-consistent solution of both equations composing Eq.(\ref{10}). Indeed putting $W_m^2=W_0^2$ we find that $\Sigma_V$ changes only due to the change of the condensate
$\tilde A_6^{4N}$ given by Eq.(\ref{9}). Direct estimation  shows that this makes less than $1$ MeV. Thus the result for the shift of $\Sigma_V$ depends on the model chosen for the description of the higher states in QCD sum rules. For example, even staying in framework of the "pole+ continuum" model, we would find another value for $\Sigma_V$, taking  different threshold values for different channels. Thus our physical result is that at $\rho=\rho_0$ the $4N$ forces increase the nucleon effective mass $m^*$ by $73$ MeV, while their contribution to $\Sigma_V$ is consistent with zero.

These are the results for the conventional value of the cutoff $\mu=882$ MeV. Since the CHPT parameter $c_1$ and the value $\delta c_1$ determined by Eq.(\ref{7b}) depend on $\mu$, the contribution of $4N$ forces also depends on $\mu$. At $\mu=600$ MeV the $4N$ interactions increase the value of the effective mass $m^*$ by $70$ MeV at $\rho=\rho_0$. At $\mu=1200$ MeV they increase $m^*$ by $75$ MeV. The value of $\Sigma_V$ changes by about $1$ MeV in these cases.

Note that our main aim was to calculate the nucleon self energies. We do not claim that our approach is precise enough to reproduce the values of the potential energy $U=\Sigma_V+\Sigma_S$ with $\Sigma_S=m^*-m$ since it is obtained as a result of subtraction of two large positive values $\Sigma_V$ and $-\Sigma_S$.

In Table 2 we demonstrate how the values of the nucleon self energies at $\rho=\rho_0$ change after separate inclusion of the nonlinear scalar condensate (NSC) and of the four-quark condensates. Note that the separate contributions are not additive, mainly because the self energies are connected with the condensates by the nonlinear expressions given by Eq.(\ref{10}).

\begin{table}
\begin{center}
\caption{Contributions of the nonlinear part of the scalar condensate (NSC) and of the four-quark vector-scalar and scalar-scalar condensates corresponding to inclusion of the $4N$ forces to the nucleon self-energies $\delta m^*$ and $\delta \Sigma_V$($\rho=\rho_0$, $\mu=882$ MeV)}.

\begin{tabular}{|c|c|c|}
\hline
    &$\delta m^*$,MeV&$\delta \Sigma_V$, MeV\\
\hline
NSC&74&8\\
\hline
VS&2&0.1\\
\hline
SS&-3&-1.2\\
\hline
\end{tabular}
\end{center}
\end{table}

\subsection{Total contribution of many body forces}

Now we compare the values of nucleon parameters with inclusion of the $3N$ and $4N$ forces with those of the gas approximation.
The two cases are shown by the solid and dotted lines in Figs.2,3 and
in the third and second lines of Table 1.

We find that the many body interactions  increase the value of the effective mass by 48 MeV and decrease that of the vector self energy by
30 MeV. Note that there is an uncertainty of about 15 MeV in the value of $m^*$ due to possible variation of the high momentum cutoff $\mu$.
It is caused mainly by the contribution to the NSC of two pion exchanges with two nucleons in the intermediate state. This uncertainty is about $1$ MeV for the vector self energy $\Sigma_V$.

We trace also the dependence of nucleon parameters on the value of $\sigma_N$. The results are shown in Table 3.
\begin{table}
\begin{center}
\caption{Dependence of nucleon parameters at $\rho=\rho_0$ on the value of nucleon sigma term $\sigma_N$}.

\begin{tabular}{|c|c|c|c|}
\hline
$\sigma_N$, MeV &$m^*$,MeV&$\Sigma_V$, MeV& $U$ MeV\\
\hline
35&746&210&27\\
\hline
40&697&201&-31\\
\hline
45&647&192&-89\\
\hline
50&597&185&-147\\
\hline
55&548&178&-202\\
\hline
60& 500&171&-257\\
\hline
\end{tabular}
\end{center}
\end{table}

The sharp dependence of nucleon effective mass $m^*$ on the value of $\sigma_N$ manifests itself in the gas approximation since the scalar condensate proportional to $\sigma_N$ determines the main part of the shift $m^*-m$.  The mass $m^*$ changes by $99$ MeV while $\sigma_N$ changes from $45$ MeV to $55$ MeV.
The many body forces make the dependence somewhat stronger mainly due to the contribution proportional to $c_1$. These two contributions exhibit linear dependence on $\sigma_N$. The $VS$ and $SS$ four quark condensates also depend on $\sigma_N$. In the latter case the dependence is not linear. However these contributions are numerically less important.

We obtained reasonable values for the nucleon potential energy for $\sigma_N=(45 \pm 5)$ MeV. At smaller values of $\sigma_N$ the potential energy runs positive. At larger values of $\sigma_N$ the value of $|U|$ is too large.

\section{Many body effects in the system of quasifree nucleons}.

Until now we treated the single pion exchange as a two-nucleon interaction. However another approach is possible. The physical nucleon in vacuum includes the self energy contribution caused by radiation and absorption of the pion. The nucleon propagator included the sum over all possible intermediate states. For the in-medium nucleon one should exclude the occupied nucleon states with momenta $p$ smaller than the Fermi momentum $p_F$-see Fig.4a. These are just the Fock terms of the nucleon interaction with one-pion exchange which can be treated as a part of the $2N$ interaction. On the other hand this contribution
can be expressed as the nucleon self energy in the matter, which can be viewed as a single particle characteristics of a quasifree nucleon.

The same refers to iterated one pion exchange \cite{13}. In the lowest order they can be expressed in terms of the fourth order self-energy diagrams for the in-medium nucleon -see Fig. 4b. We define the sigma term $\hat\sigma_N$ for quasifree nucleon by  Eq.(\ref{10a}).
We present
\begin{equation}
\hat\sigma_N(\rho)=\sigma_N+m_{\pi}^2f_1(\rho),
\label{20}
\end{equation}
with $f_1(\rho)$ the sum of the right hand sides of Eqs.(4), (6), (7), (8) and (9) of \cite{9}. Also the matrix element of the operator $\sum_i\bar q^iq^i$ for the quasifree nucleon is
$$\hat \kappa_N=\kappa_N+\frac{2m_{\pi}^2}{m_u+m_d}f_1(\rho).$$
We find  $\hat\sigma_N=63$ MeV at $\rho=\rho_0$.
This provides $m^*=433$ MeV which is much smaller than in the gas approximation for free nucleons.

Now we include the $2N$ interactions between the quasifree nucleons of the matter.  They determine the $3N$ interactions between the probe nucleon and the matter in the quasifree nucleon presentation. We calculate the contribution to the NSC
\begin{equation}
\hat\sigma^{(2)}=m_{\pi}^2f_2(\rho),
\label{21}
\end{equation}
with $f_2(\rho)$ the sum of the right hand sides of Eqs.(11), (14), (17), (18) and (20) of \cite{9}. We must include the contribution of the four-quark condensates found in \cite {5} with $\kappa_N$ changed to $\hat\kappa_N$ in Eqs. (33) and (36). We find $\hat\sigma^{(2)}=-23 $MeV at $\rho=\rho_0$.
The main contribution comes from two-pion exchanges with virtual $\Delta$ excitations.

To include the $3N$ interactions between the quasifree nucleons of the matter which determine the $4N$ interactions between the probe nucleon and the matter in the quasifree nucleon presentation, we find the corresponding NSC
\begin{equation}
\hat\sigma^{(3)}=m_{\pi}^2f_3(\rho)+\delta\hat \sigma^{(3)}
\label{22}
\end{equation}
Here $f_3(\rho)$ is the sum of the right hand sides of Eqs.(12), (13), (22) and (23) of \cite{9}, while $\delta\hat \sigma^{(3)}$ is determined by Eq.
(\ref{7a}) with $\sigma^{(2)}$ replaced by $\hat \sigma^{(2)}$. We obtain $\hat\sigma^{(3)}=-5.3 $MeV at $\rho=\rho_0$.
The leading contributions are determined by the two-pion exchanges with virtual $\Delta$ excitations and by the term $\delta\hat \sigma^{(3)}$.

Since $\hat \sigma_N >|\hat \sigma^{(2)}| >|\hat \sigma^{(3)}|$ we obtain the strict hierarchy of contributions of the many body forces to the vector self energy $\Sigma_V$, to the effective mass $m^*$ and to the potential energy $U$.

Density dependence of the nucleon characteristics with inclusion of the $2N$,  $3N$ and $4N$ interactions between the probe nucleon and the  matter in the quasifree nucleon presentation is shown in Fig.5. The density dependence of the residue $\lambda_m^2$ and of the continuum threshold $W_m^2$ are shown in Fig.6. More detailed data in the vicinity of the point $\rho=\rho_0$ are presented in Table 4.

Since $|\hat \sigma^{(2)}| \gg \sigma^{(2)}$ the role of the four-quark condensates increases in the quasifree nucleons presentation. The contribution of the $4N$ forces to the vector self energy $\Sigma_V$ making 9 MeV at $\rho=\rho_0$ is mainly due to the SS four-quark condensate.

\begin{table}
\caption{The density dependence of nucleon parameters . For each value of $\rho/\rho_0$
the upper line shows the result for the gas of the quasifree nucleons. In the second and third lines the $3N$ and $4N$ forces between the probe nucleon and the matter are added ($\mu=882$ MeV).}
\begin{center}
\begin{tabular}{|c|c|c|c|c|c|}
\hline
$\rho/\rho_0$&$m^*$,MeV&$\Sigma_V$, MeV&$U$, MeV&$\lambda_m^2$,GeV$^6$&$W_m^2$, GeV$^2$\\
\hline
0.90&509&182&-237&1.14&1.60\\
&646&167&-116&1.55&1.83\\
&664&173&-92&1.57&1.85\\
\hline
0.95&472&195&-262&1.10&1.60\\
&627&175&-127&1.53&1.83\\
&648&182&-98&1.54&1.85\\
\hline
1.00&433&207&-288&1.07&1.59\\
&607&183&-138&1.52&1.83\\
&631&192&-105&1.52&1.84\\
\hline
1.05&392&221&-316&1.04&1.59\\
&587&191&-150&1.50&1.83\\
&613&202&-112&1.50&1.84\\
\hline
1.10&347&235&-347&1.02&1.59\\
&567&199&-162&1.49&1.83\\
&595&212&-129&1.48&1.85\\
\hline

\end{tabular}
\end{center}
\end{table}

Note that the results for the total contribution of the $2N$ and $3N$ interactions are close to those obtained in the previous Section but are not identical to them. The main reason is that now we replace $\sigma_N$ by $\hat\sigma_N$ in expression for $c_1$. This changes the contribution of the right hand side of Eq.(20) of \cite{9} to the function $f_2(\rho)$ (Eq.(\ref{20})) and of those of the right hand sides of Eqs.(22,23) to the function $f_3(\rho)$. We replaced also $\sigma^{(2)}$ by $\hat \sigma^{(2)}$ in Eq.(\ref{7b}) for $\delta c_1$. Thus the values of $m^*$ at $\rho=\rho_0$ differ by 16 MeV.

\section{Summary}

We calculated  the contribution of many body forces to the nucleon characteristics in symmetric nuclear matter. We found the contribution of the $3N$ interactions earlier \cite{5}. Now we add that of the $4N$ forces.

The main contribution of the $4N$ forces to the effective mass $m^*$ is determined by the nonlinear scalar condensate with inclusion the $3N$ interactions between the nucleons of the matter. The contribution to the vector self energy of the nucleon is dominated by the four-quark condensates in the scalar-scalar and
 vector-scalar channels with two diquark operators acting on two different nucleons of the matter. The values of these condensates are determined by the nonlinear scalar condensate with inclusion of the $2N$ interactions between the nucleons of the matter. Connection between our approach and the picture based on the nucleon-nucleon interactions is illustrated by  Fig.1.

The results of calculations at $\sigma_N=45$ MeV are presented in Figs.2,3 and in Table 1. We found that at $\rho=\rho_0$ the $4N$ forces increase the effective mass $m^*$ by 73 MeV. Earlier we obtained that inclusion of $3N$ forces diminishes the value of $m^*$ by 26 MeV \cite{5}. Thus the total contribution of the $3N$ and $4N$ forces increase the gas approximation value by $47$ MeV. The contribution of $4N$ forces is larger than that of the $3N$ forces. The $4N$ forces change the value of the vector self energy by several MeV while the $3N$ forces diminished it by $37$ MeV. Thus the contribution of the $4N$ forces to the vector self energy is much smaller than that of the $3N$ forces.

We traced the dependence of the nucleon parameters on the actual value of $\sigma_N$. We found reasonable values for the nucleon potential energy for $\sigma_N=(45 \pm 5)$ MeV. At larger values of $\sigma_N$ the bound of the nucleons in the matter looks to be too strong.

Calculations of the nucleon energy in the chiral effective field theory reviewed in \cite{14} demonstrated that the $4N$ forces give smaller contribution
than the $3N$ forces. Our results can not be directly compared with those of \cite{14}. In \cite{14} both Hartree and Fock terms in the scalar channel have been obtained. Our analysis corresponds rather to the mean field approach. The Fiertz transformed Fock terms of \cite{14} contribute to both scalar and vector channels in our approach. In calculations carried out in framework of traditional nuclear physics the authors of \cite{15} found that the
contribution of the $4N$ forces to the binding energy of the nucleon in nuclear matter is smaller than that of the $3N$ forces. On the other hand, a number of other calculations showed the $3N$ and $4N$ correlations to be of the same order \cite{16}.

We developed also another approach in which the one-pion exchanges and the iterated one-pion exchanges which can be expressed in terms of the nucleon self energy are treated as the single-nucleon characteristics. In this approach the gas approximation corresponds to system of noninteracting nucleons with their pion clouds (quasifree nucleons). Parameters of the probe nucleon calculated in the gas approximation for the quasifree nucleons corresponds to inclusion of the $2N$ forces. Density dependence of the nucleon characteristics corresponding to inclusion of the $2N$ , $3N$ and $4N$ interactions between the probe nucleon and the  nucleons of the matter in the quasifree presentation is given in Fig.5,6  and in Table 4. The many  body interactions between the probe nucleon and the nucleons of the matter respect the strict hierarchy. The contribution of $3N$ interactions to the effective mass $m^*$, to the vector self-energy $\Sigma_V$ and to the potential energy $U$ are much smaller than those of $2N$ interactions. The contributions of $4N$ forces are much smaller than those of the $3N$ forces.

{}

\section{Figure captions}

\noindent
Fig.~1. Feynman diagrams of nucleon-nucleon interactions corresponding to inclusion of the $4N$ forces. Solid lines are for the probe nucleon, bold lines denote the nucleons of the matter. The shaded block denotes the pion field. The dashed line denotes the scalar meson. The dashed-dotted line  denotes the scalar or vector meson.\\

\noindent
Fig.~2. Density dependence of the effective nucleon mass $m^*$, of the vector self energy $\Sigma_V$ and of the single-particle potential energy $U$. The horizontal axis corresponds to the density $\rho$ related to its saturation value $\rho_0$. The vertical axis is for $m^*$, $\Sigma_V$ and $U$. The dotted lines-only the $2N$ forces between the nucleons of the matter are included. Dashed lines-the $2N$ and $3N$ forces are included. Solid lines-the $2N$, $3N$ and $4N$ forces are included.\\

\noindent
Fig.~3. Density dependence of the nucleon residue $\lambda_m^2$ and of the continuum threshold $W_m^2$.
The notations are the same as in Fig.~2.\\

\noindent
Fig.~4. The Feynman self energy diagrams for the nucleon in the matter, corresponding to  the one-pion exchange (Fig.a) and to the iterated one-pion exchange (Fig.b). Solid lines are for nucleons, wavy lines are for the pions. The cross on the line marks that the nucleon is on the mass shell.\\

\noindent
Fig.~5. Density dependence of the effective nucleon mass $m^*$, of the vector self energy $\Sigma_V$ and of the single-particle potential energy $U$
in the quasufree nucleons presentation.
The notations are the same as in Fig.2.\\

\noindent
Fig.~6 Density dependence of the nucleon residue $\lambda_m^2$ and of the continuum threshold $W_m^2$ in the quasifree nucleons presentation.
The notations are the same as in Fig.~2.\\

\newpage

\begin{figure}
\centerline{\epsfig{file= 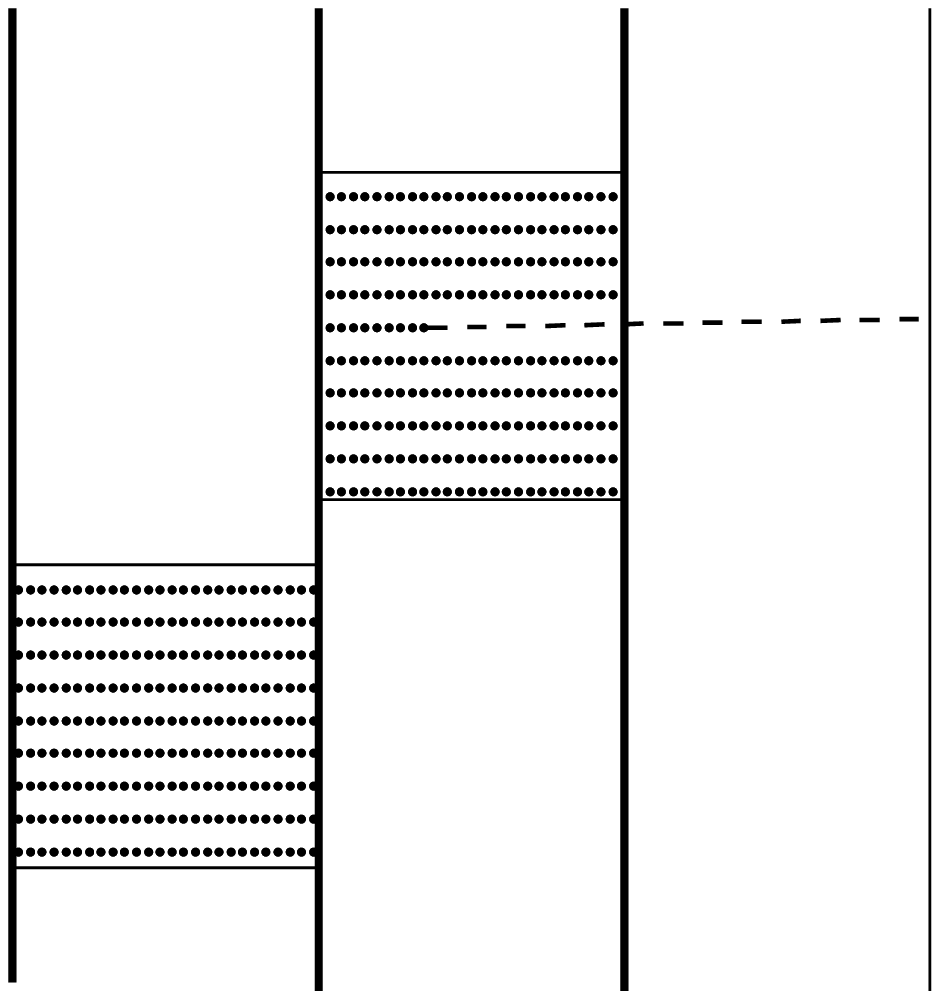,width=5.5cm} \hspace{2cm} \epsfig{file= 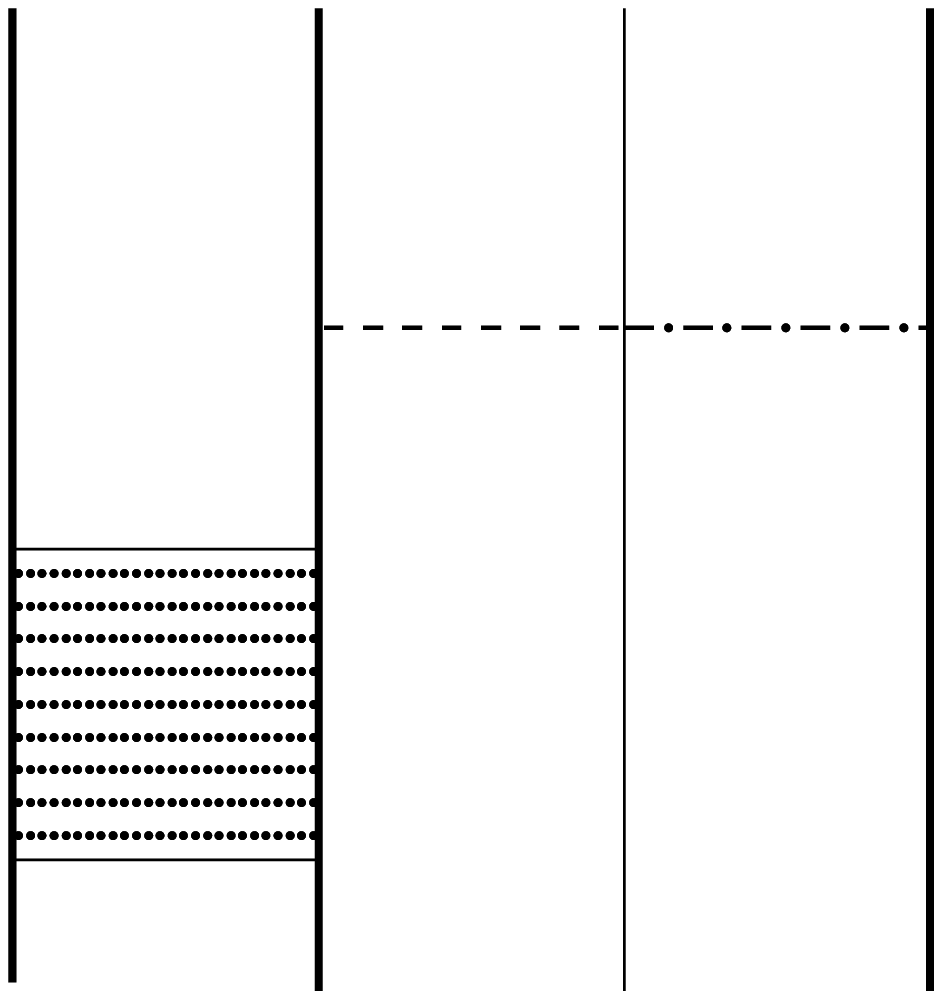,width=5.5cm}}
\caption{}
\end{figure}

\begin{figure}
\centerline{\epsfig{file= 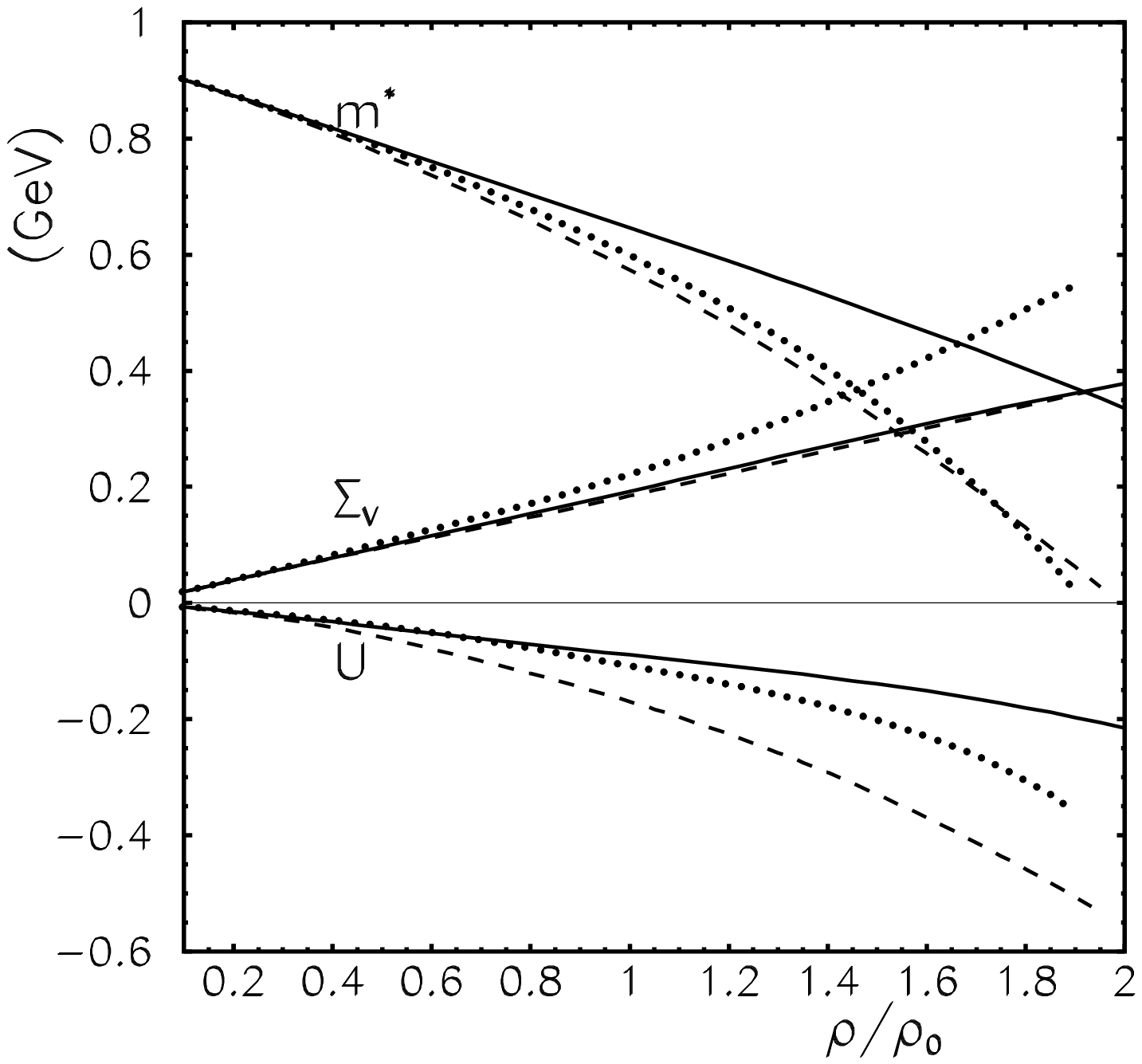,width=8.0cm}}
\caption{}
\end{figure}

\begin{figure}
\centerline{\epsfig{file= 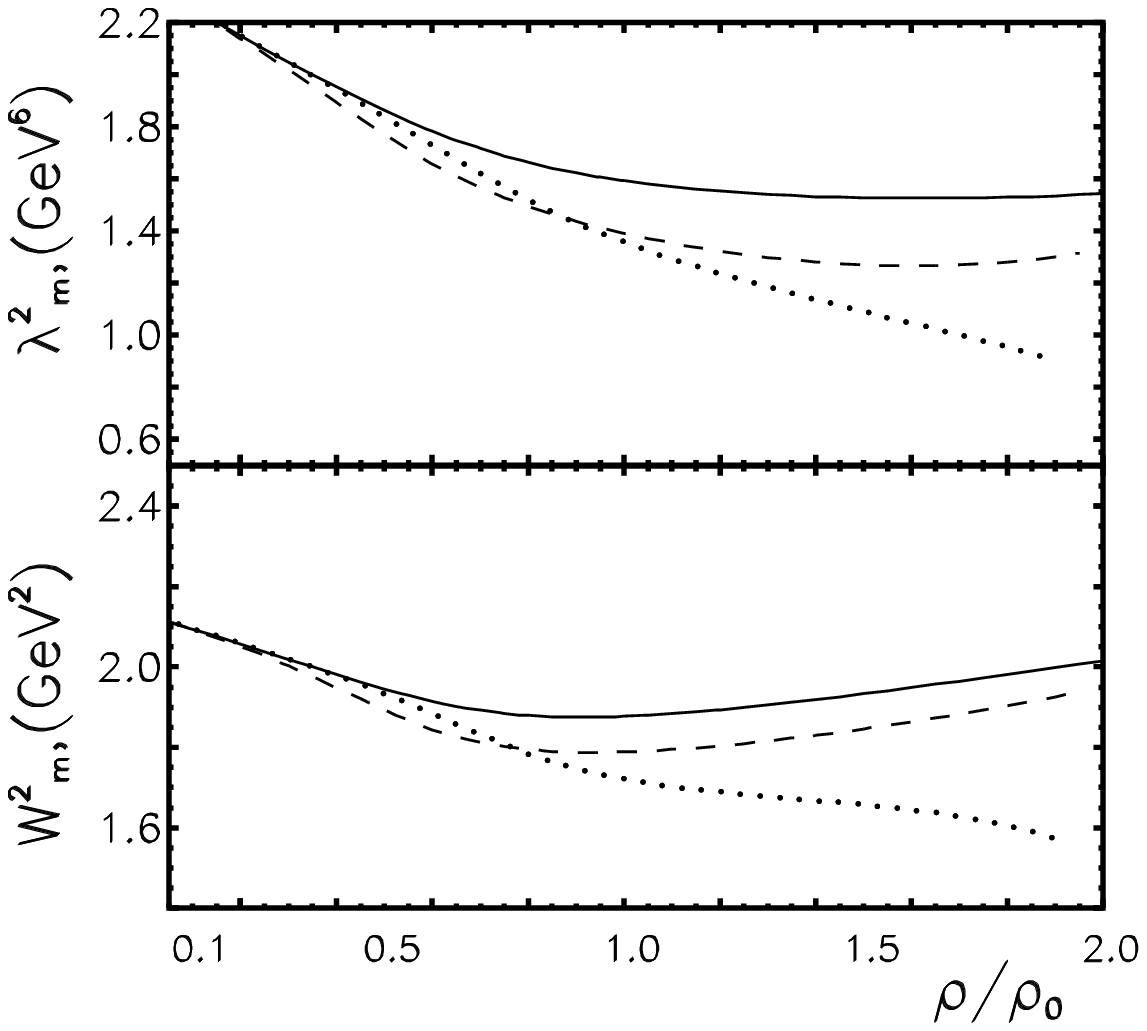,width=8.5cm}}
\caption{}
\end{figure}

\begin{figure}

\vspace{-1.5cm}

\centerline{\epsfig{file= 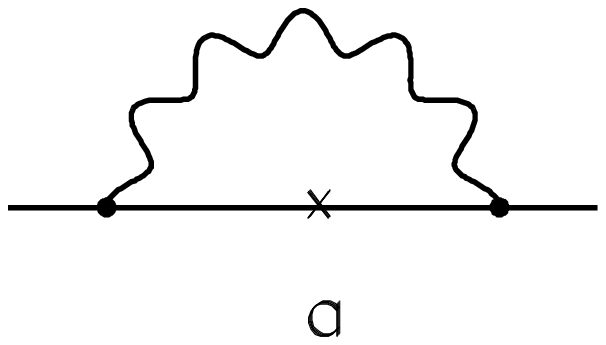,width=6.cm}}

\vspace{-1.0cm}

\centerline{\epsfig{file= 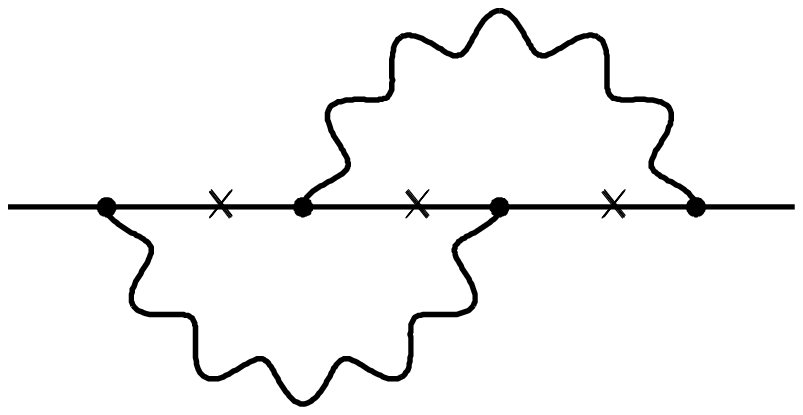,width=6.0cm}\hspace{-0.7cm}
\epsfig{file= 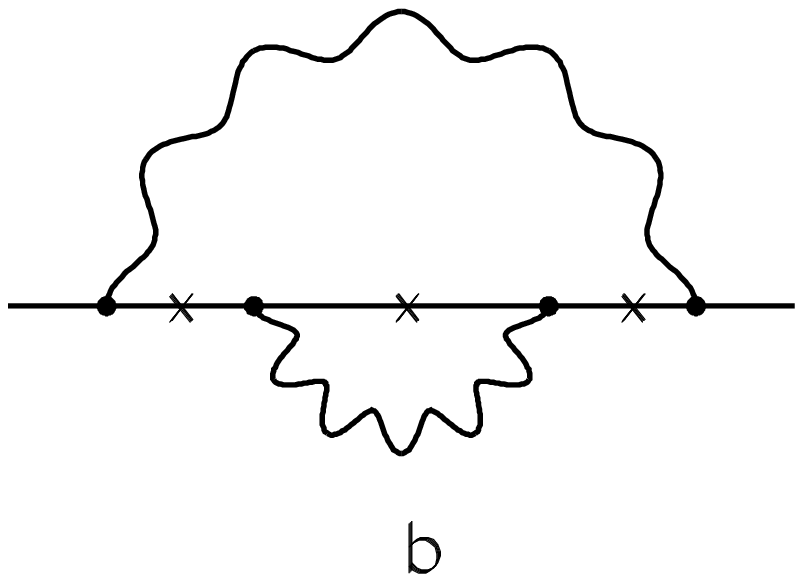,width=6.0cm} \hspace{-0.7cm}\epsfig{file= 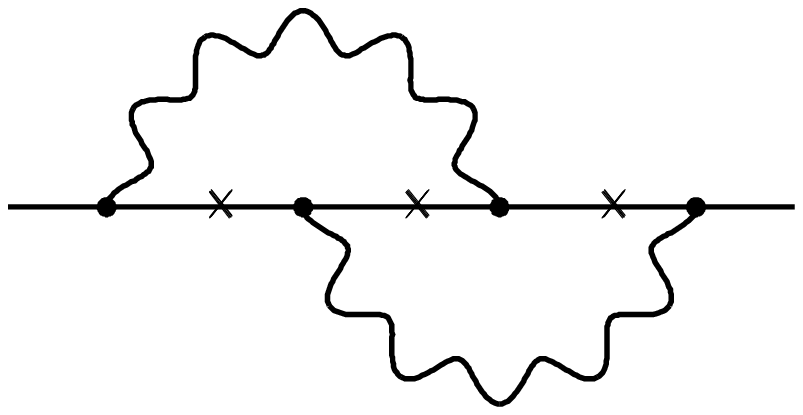,width=6.0cm}}
\caption{}
\end{figure}

\begin{figure}
\centerline{\epsfig{file= 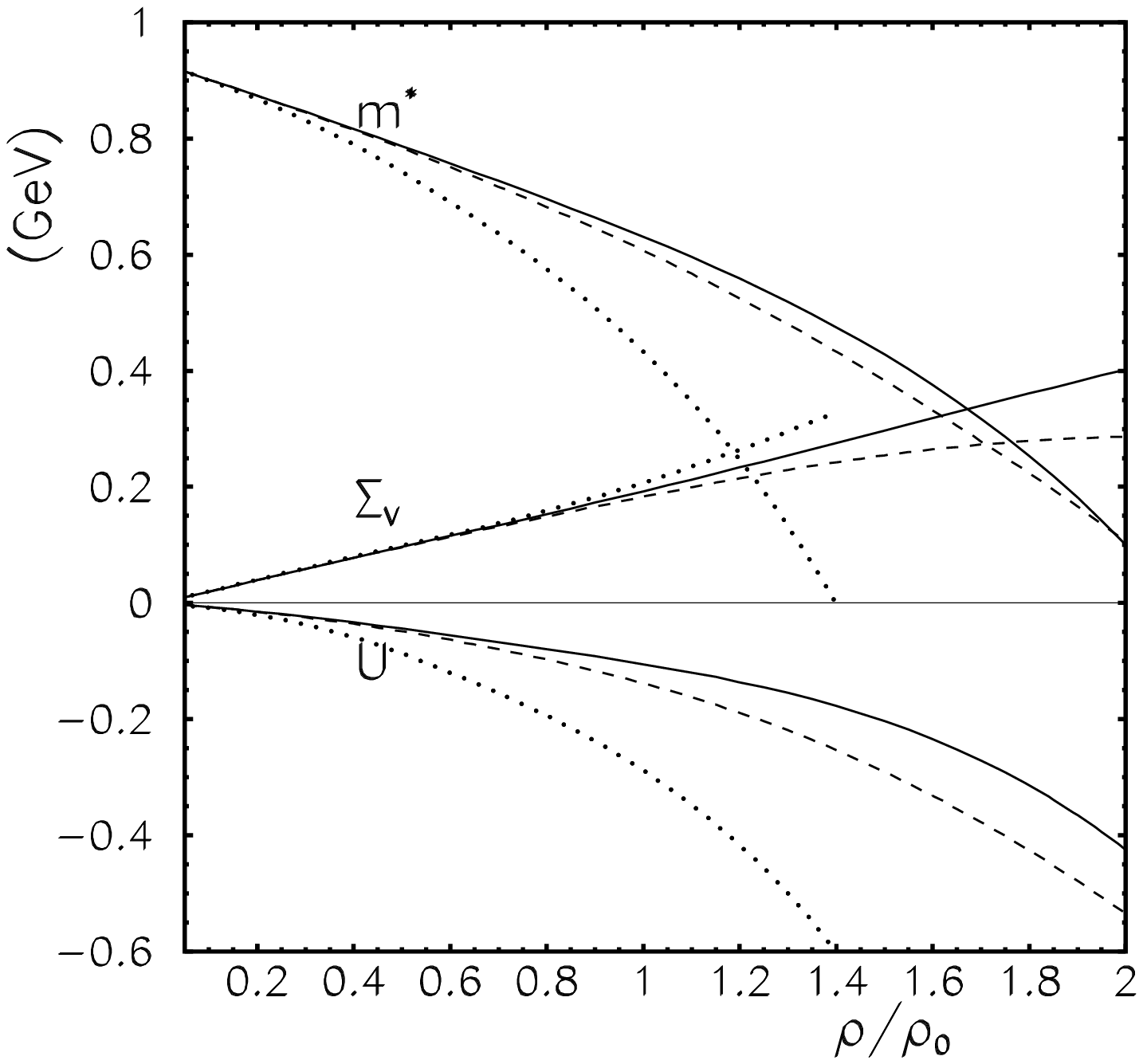,width=8.0cm}}
\caption{}
\end{figure}

\begin{figure}
\centerline{\epsfig{file= 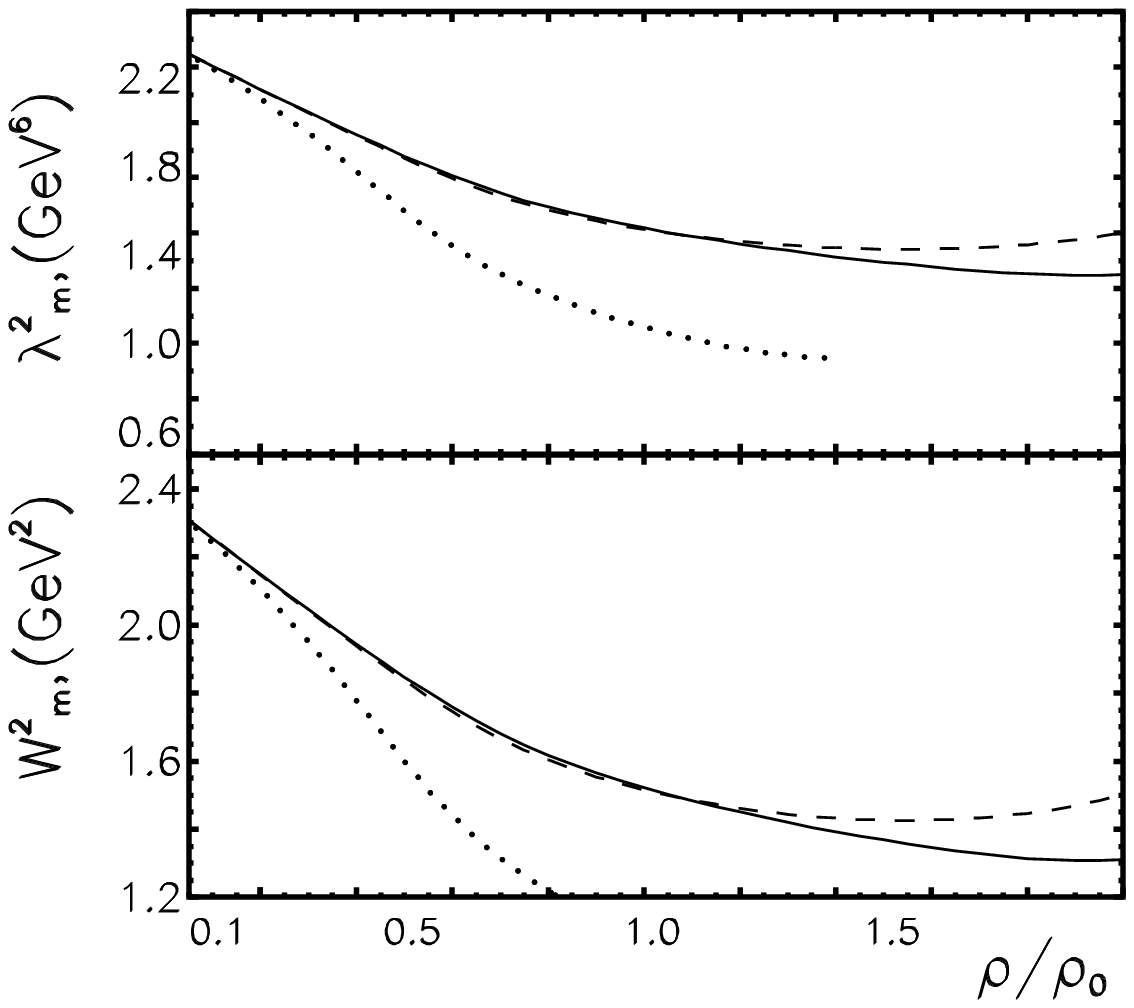,width=8.5cm}}
\caption{}
\end{figure}

\end{document}